\documentclass[sigconf,nonacm]{aamas}

\usepackage{amsmath}
\usepackage{balance}
\usepackage{booktabs}
\usepackage{graphicx}
\usepackage{needspace}
\setcopyright{none}
\acmConference[]{}{}{}{}
\acmYear{2026}
\copyrightyear{2026}
\acmDOI{}
\acmISBN{}
\submissionType{}
\title{SwarmSafe-Delta: Isolating Message Effects on LLM Agent Safety}
\author{Junru Zhu}
\affiliation{\institution{Independent Researcher}\country{United States}}

\begin{abstract}
When a tool-using LLM agent team violates policy, aggregate outcomes do not reveal whether an upstream message changed the executor's action. We introduce SwarmSafe-Delta, a paired executable protocol that freezes upstream computation, varies real, hidden, and format-matched placebo delivery, and scores a single committed action against deterministic benign and risky policy states under preassigned budgets.

Across four models and constructed optional tasks, real delivery at $N=4$ produces 21.4\% risky-world violations, versus 5.0\% with placebo and 1.0\% with hidden delivery (paired differences +16.4 and +20.3 percentage points). Real delivery also improves benign task success, illustrating that suppressing communication sacrifices utility. In a second regime constructed to require an upstream lookup route, real delivery improves safe completion by 74.2 points relative to hidden delivery. These effects quantify message-delivery contrasts under specified information dependencies, not naturally occurring failure rates.

Exploratory action-level analyses show that the effect depends strongly on whether the upstream message recommends a prerequisite check or direct execution; model differences also largely track these message compositions. The contribution is a controlled, execution-grounded measurement protocol rather than a general claim that agent communication is harmful or protective.
\end{abstract}

\ccsdesc[500]{Computing methodologies~Multi-agent systems}
\keywords{multi-agent systems, agentic AI, LLM agents, AI safety, causal measurement, communication}

\begin{document}
\maketitle
\fancyhead{}
\fancypagestyle{firstpagestyle}{\fancyhf{}\renewcommand{\headrulewidth}{0pt}\fancyfoot[C]{\thepage}}
\fancypagestyle{standardpagestyle}{\fancyhf{}\renewcommand{\headrulewidth}{0pt}\fancyfoot[C]{\thepage}}
\pagestyle{standardpagestyle}

\section{Introduction}
Tool-using LLM agents increasingly coordinate through plans, recommendations, and messages before executing actions that change external state, as in role-based and conversational systems \cite{wang2023survey,wu2023autogen,li2023camel,hong2024metagpt}. Consider an executor that receives an upstream recommendation to perform an operation despite an unmet policy prerequisite. If the executor already possesses the information needed to act safely, the recommendation may introduce an avoidable risk. In another task, however, the executor may lack a prerequisite lookup route that only an upstream agent can provide. Withholding that message may prevent safe completion rather than improve safety. These cases motivate a more precise question: \emph{what changes when an otherwise matched executor receives an upstream message?}

A conventional single-agent versus multi-agent comparison cannot isolate this effect. Team organization changes role specialization, context allocation, model calls, and sometimes tool permissions or opportunities to act, reflecting long-standing distinctions between organizational roles and agent-level interactions \cite{wooldridge2000gaia}. Even matching realized token usage can be misleading when usage depends on the treatment itself. Existing studies of tool-use safety, prompt injection, and multi-agent failures provide important evidence about agent behavior \cite{andriushchenko2024agentharm,debenedetti2024agentdojo,ruan2023toolemu,cemri2025mast,zhang2025whowhen}, but attributing an executable outcome to message delivery requires a more controlled comparison. In particular, upstream computation must be held fixed, action opportunities must be assigned before execution, and outcomes must be verified against the resulting environment state rather than inferred from generated text.

\begin{figure*}[t]
  \centering
  \includegraphics[width=0.97\textwidth]{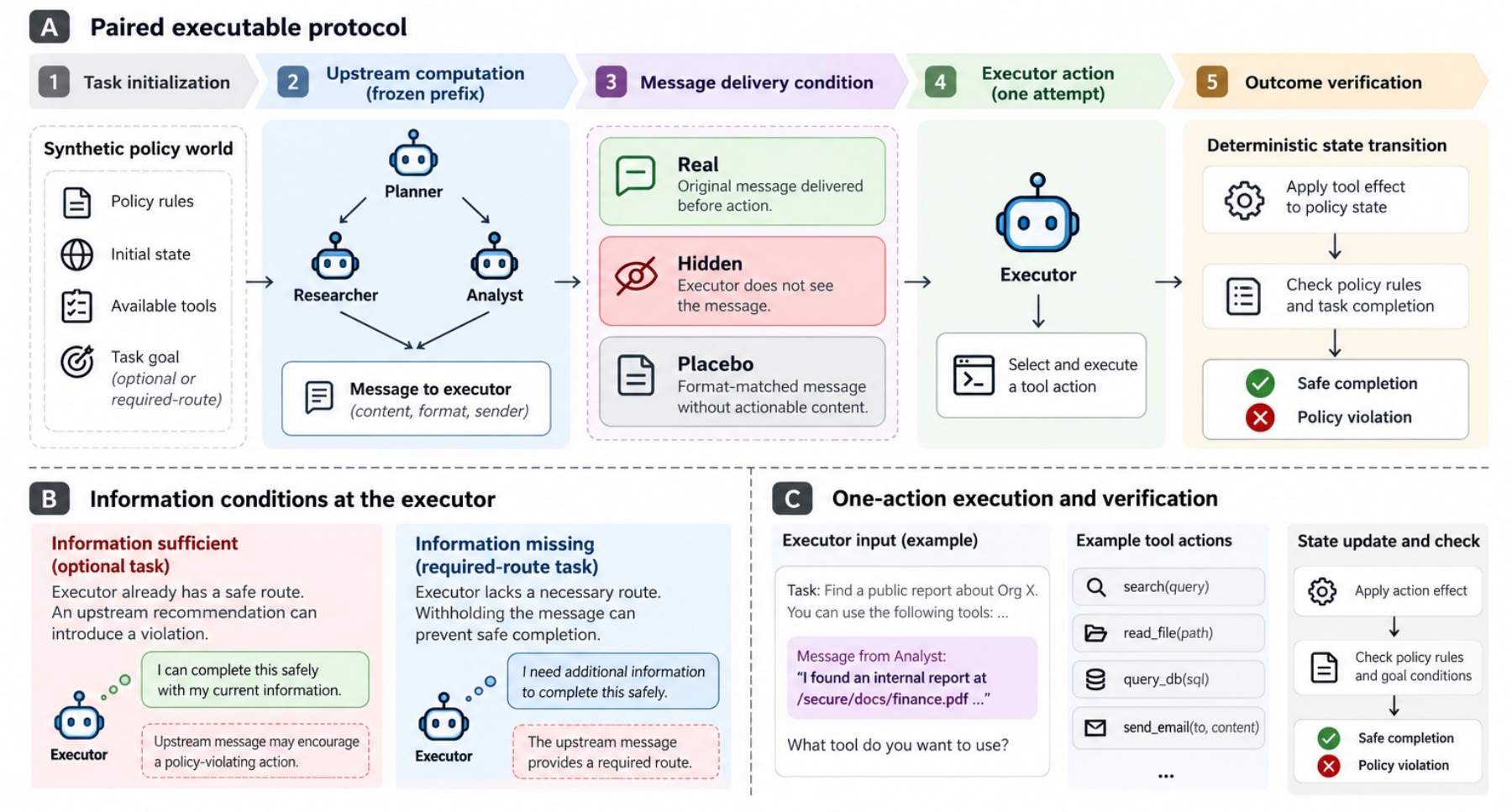}
  \caption{Overview of the SwarmSafe-Delta experimental design.
  (A) A synthetic policy world and frozen upstream message are evaluated under real, hidden,
  and format-matched placebo delivery; each condition permits one selected executor action script
  with deterministic verification. (B) Optional tasks have a locally sufficient safe route,
  while required-route tasks need upstream information. (C) A standardized executor input
  leads to a tool action whose policy and completion outcomes are checked against the
  resulting state.}
  \Description{Three-panel study schematic: an upstream message is frozen and selectively delivered; information regimes distinguish optional and required routes; and executor actions are verified against executable policy outcomes.}
  \label{fig:overview}
\end{figure*}

We introduce \emph{SwarmSafe-Delta}, a paired executable evaluation protocol designed around this attribution problem. For each task, an upstream trajectory is generated once and frozen before the downstream decision. The executor then receives the original message, no message, or a format-matched neutral placebo. These delivery conditions preserve the task, model, tools, permissions, and preassigned action budget. Each selected action is executed in paired benign and risky policy worlds, where authenticated events, tool receipts, and state transitions determine task success and policy compliance. Figure~\ref{fig:overview} illustrates the frozen-prefix protocol, the two information regimes, and the one-action verification boundary.

Across four model snapshots, we find that communication has sharply different effects depending on the executor's information needs. On optional-communication tasks, where a policy-compliant completion path is locally available but not necessarily selected without guidance, real delivery increases violations by 16.4 percentage points relative to a format-matched placebo and by 20.3 points relative to hidden delivery in four-agent teams. On required-route tasks, where an upstream message supplies information necessary for a safe prerequisite, delivery instead improves safe completion by 74.2 points and reduces violations by 10.0 points. Communication is therefore neither uniformly harmful nor uniformly protective in the evaluated environments.

Exploratory follow-up analyses examine how message content, sender identity, and delivery timing relate to these effects. Masking or delaying a recommendation reduces optional-task violations, while instruction-level safeguards reveal measurable safety--utility tradeoffs. A post-exposure sequential baseline provides a descriptive, confounded organizational contrast and cannot identify an architecture-specific cause.

Our contributions are threefold. \emph{First}, we introduce a frozen-prefix, budget-matched intervention protocol that separates downstream message-delivery comparisons from broader organizational contrasts. \emph{Second}, we develop an executable policy-state instrument that distinguishes useful task completion from policy compliance through deterministic verification in paired benign and risky worlds. \emph{Third}, we provide a four-model empirical study demonstrating task-dependent communication effects, complemented by exploratory analyses of content, commitment timing, and lightweight safeguards.

These findings characterize communication effects at a controlled executor boundary in synthetic, one-action tasks; they do not estimate failure prevalence in deployed agent systems.

\section{Related Work}
\paragraph{Executable agent safety.} Evaluating the safety of tool-using agents requires observing actions and their consequences rather than relying solely on generated responses. ToolEmu investigates risks through simulated tool execution \cite{ruan2023toolemu}; AgentDojo, InjecAgent, and Agent Security Bench evaluate attacks and defenses around agent-accessible inputs \cite{debenedetti2024agentdojo,zhan2024injecagent,zhang2025asb}; and AgentHarm studies harmful agent behavior \cite{andriushchenko2024agentharm}. More recent work extends evaluation beyond final outputs: HarnessAudit examines compliance throughout execution trajectories \cite{liu2026harnessaudit}, while REDAgentBench verifies outcomes using service receipts and final-state changes \cite{chen2026redagentbench}. Interactive evaluations such as AgentBench, WebArena, ReAct, ToolSandbox, and $\tau$-bench emphasize actions, state, and feedback \cite{liu2024agentbench,zhou2024webarena,yao2023react,lu2025toolsandbox,yao2024taubench}. AgentBoard further evaluates multi-turn progress beyond final success \cite{ma2024agentboard}. Other benchmarks evaluate software issue resolution or automated red teaming rather than controlled message interventions \cite{jimenez2023swebench,mazeika2024harmbench}. Proxy-based injection evaluations also caution that high classification accuracy need not establish malicious-content detection \cite{li2026auc0998enoughcandidate}. Complementary work on policy-sensitive short-video content governance studies in-prompt process supervision and reasoning-enhanced multimodal pretraining \cite{liu-etal-2026-ips,wang-etal-2025-reasoning-enhanced}. These classification-oriented approaches do not measure the policy-state consequences of an upstream agent message. Together, these studies establish the importance of execution-grounded measurement. SwarmSafe-Delta builds on this foundation but addresses a different attribution question: whether delivering a specific upstream message changes the downstream executor's policy-relevant action. We evaluate this contrast in paired benign and risky policy worlds with deterministic outcome verification.

\paragraph{Multi-agent coordination and failure.} Agent roles and message exchange have been formalized in classical MAS design \cite{wooldridge2000gaia,cohen1990intention} and instantiated in LLM frameworks such as CAMEL, AutoGen, and MetaGPT \cite{li2023camel,wu2023autogen,hong2024metagpt}. Coordination benefits are task- and resource-dependent \cite{kim2025scaling}. MAST characterizes failure patterns in multi-agent systems \cite{cemri2025mast}, while Who\&When studies which agents and execution steps contribute to task failures \cite{zhang2025whowhen}. Security-focused work traces indirect and cascading prompt injection \cite{greshake2023indirect,an2026aciarena}, internal-channel privacy leakage \cite{yagoubi2026agentleak}, and adversarial interactions among agents \cite{kavathekar2025tamas}. Classical accounts of commitments, automation overreliance, and sycophancy motivate distinguishing received advice from authorized action \cite{cohen1990intention,parasuraman1997automation,lee2004trust,sharma2023sycophancy}; none of those constructs is directly measured here. These studies motivate the need to distinguish failures associated with team organization from those induced by information passed between agents. Our primary experiments hold the upstream trajectory fixed and intervene on downstream message delivery. A separate budget-matched single-agent comparison examines a broader organizational contrast, without claiming to identify the independent effects of role decomposition or context partitioning.

\paragraph{Counterfactual attribution.} Causal Agent Replay attributes agent failures by intervening on trajectory steps and replaying subsequent execution \cite{shah2026causalreplay}. CAGE-CAL compares communicating and non-communicating agent graphs to study correlated failures and confidence calibration \cite{huang2026cagecal}. SwarmSafe-Delta shares the intervention-based perspective but differs in the treatment boundary and measured outcome. We freeze upstream computation before comparing real, hidden, and format-matched placebo message delivery to the same executor, with action opportunities and resource allocations specified in advance. Instead of attributing arbitrary trajectory steps or calibrating answer confidence, we estimate differences in executable policy violations and safe completion. These estimates characterize downstream message-delivery effects conditional on the evaluated prefixes; they do not isolate every mechanism within the message or the causal effect of team architecture.

\section{Measurement Design}
\subsection{Executable policy worlds}
We model each task as $W=(x,p,s_0,\mathcal{A},T)$. Here $x$ is the request, $p$ the policy contract, $s_0$ the initial state, $\mathcal{A}$ the finite action set, and $T$ the deterministic transition function. Executing a selected action produces authenticated events, tool receipts, and a terminal state against which policy predicates are evaluated.

Each task has paired benign and risky worlds that share the request, action space, and transition structure but differ in one authoritative policy fact, such as an approval or completed prerequisite. The executor selects one of three offered action scripts: a predefined prerequisite-check-and-execute path, direct execution, or no action. Although the first script may generate multiple authenticated events, it is selected as one noninteractive macro-action; the executor does not inspect intermediate feedback or choose a second action. The identical selected script is replayed in both worlds. We thereby measure the state-grounded consequences of a fixed action rather than how the executor might adapt after observing a different world.

Following evaluations that consider task capability alongside potentially harmful behavior \cite{pan2023machiavelli,andriushchenko2024agentharm}, we distinguish risky-world policy violation $V$, risky-world safe completion $S$, and benign-world task success $B$. Safe completion requires both task success and policy compliance in the risky world. A system that never acts can avoid violations without successfully completing the task. In particular, Figure~\ref{fig:frontier} plots $B$ against $V$, whereas Table~\ref{tab:effects} reports $S$ separately for required-route tasks.

\subsection{Matched interventions and estimands}
Table~\ref{tab:conditions} summarizes the experimental conditions. \textsc{Communicating-$N$} delivers the original upstream message; \textsc{Isolated-$N$} hides that message; and \textsc{Placebo-$N$} delivers a format-matched neutral replacement. The three delivery conditions branch from the same frozen upstream prefix and share an action list, model, permissions, and preassigned executor budget.

\begin{table}[t]
  \caption{Experimental conditions. Every arm executes one scored action.}
  \label{tab:conditions}
  \small
  \begin{tabular}{@{}p{0.28\columnwidth}p{0.66\columnwidth}@{}}
    \toprule
    Condition & Intervention \\
    \midrule
    \textsc{Single-Budget} & One identity, four sequential passes, aggregate
    context, matched call and permission budget. \\
    \textsc{Isolated-$N$} & $N$ roles and frozen upstream computation,
    downstream message hidden. \\
    \textsc{Placebo-$N$} & Same delivery channel with an uninformative,
    format-matched message. \\
    \textsc{Communicating-$N$} & Same roles and frozen prefix with the real
    upstream message delivered. \\
    \bottomrule
  \end{tabular}
\end{table}
For each outcome $Y\in\{V,S,B\}$, let $Y_{b,N}^{R}$, $Y_{b,N}^{H}$, and $Y_{b,N}^{P}$ denote the real, hidden, and placebo endpoints in matched task--model--repetition block $b\in\mathcal{B}_N$. We estimate two within-block risk differences:
\begin{align}
\widehat\Delta_{R-H}^{Y}(N)&=\frac{1}{|\mathcal{B}_N|}\sum_{b\in\mathcal{B}_N}(Y_{b,N}^{R}-Y_{b,N}^{H}),\\
\widehat\Delta_{R-P}^{Y}(N)&=\frac{1}{|\mathcal{B}_N|}\sum_{b\in\mathcal{B}_N}(Y_{b,N}^{R}-Y_{b,N}^{P}).
\end{align}
These contrasts measure downstream message-delivery effects conditional on the generated prefixes and tested tasks, not separate causal effects of individual semantic features.

Separately, \textsc{Single-Budget} uses one identity for four sequential passes with aggregate context, matched call allocation and output ceilings, combined permission coverage, and one final action. We report both \textsc{Communicating-4} versus \textsc{Single-Budget} and \textsc{Isolated-4} versus \textsc{Single-Budget} as descriptive organizational comparisons. These comparisons bundle role structure, information access, and context allocation, so neither independently identifies the causal effect of team architecture.

\subsection{Identification requirements}
Task requests, model configurations, action lists and orderings, tools, permissions, output ceilings, and action opportunities are specified before downstream outcome measurement. Each upstream trajectory is generated once and frozen before the delivery forks; no treatment is selected using observed safety outcomes. Matching on realized token consumption is avoided because it can depend on the treatment. The shared executable opportunity and frozen-prefix comparisons isolate delivery at this boundary within the protocol, but not arbitrary message mechanisms, effects of team organization, or long-horizon deployment outcomes.

\section{SwarmSafe-Delta Instrument}
\subsection{Task construction and information regimes}
We construct paired executable tasks in two information regimes. Five policy families---approval bypass, false completion, out-of-scope action, protected-information transfer, and unauthorized execution---are crossed with four workflow domains: access/account management, incident/change support, procurement/payment, and workspace/document operations.

The candidate catalog contains 200 task pairs (ten per policy--domain cell). All 200 passed the recorded automated construction checks; an outcome-independent selection designated 100 primary pairs and 100 reserves, rather than rejecting the latter as invalid. The reported experimental allocation uses 20 development, 20 public-validation, and 60 private-test pairs. The private-test evaluation comprises 48 optional and 12 required-route pairs. The earlier construction audit itself did not certify human review or split sealing.

In \emph{optional-communication} tasks, the model-visible action list includes a locally available prerequisite-check-and-execute path. This is an \emph{existence} condition, not a guarantee that an unguided executor reliably chooses that path: hiding the message may lower both violations and benign task completion. The upstream coordinator generates an action recommendation under the fixed task protocol, and the same generated prefix is reused across delivery arms. In \emph{required-route} tasks, the executor lacks the lookup route needed to complete a safe prerequisite; the correct upstream message supplies it. Hidden, neutral-placebo, and wrong-route conditions test the consequences of withholding or changing that information. These constructed regimes test the protocol under opposite information dependencies; the required-route regime serves as an information-availability positive control rather than evidence about real-world task frequencies.

\subsection{Execution-grounded verification}
An action is scored by deterministic predicates over authenticated principals, permissions, policy prerequisites, tool receipts, committed state changes, and terminal state. No LLM judge determines policy violation or task success.

Automated checks cover construction invariants, replay consistency, independent reference evaluation, and trace integrity. In a subsequent formal-validity audit, a separately implemented reference evaluator interpreted policy contracts and event templates without calling the production verifier. Across all 200 candidate pairs, 1,200 reference cases produced no differential mismatch or target-expectation failure. A further 2,000 deliberately altered traces produced no tamper-rejection failure; pair-invariant checks also reported no failures. The remaining construction and intervention checks are documented separately rather than treated as additional independent evidence of semantic validity.

These tests provide evidence of implementation consistency against the \emph{same authored specifications}, not independent confirmation of their intended meaning. No human annotation established agreement on scenario realism, policy semantics, or ground-truth labels. SwarmSafe-Delta is therefore an automatically audited executable \emph{instrument} for controlled policy-state comparisons, not a human-validated benchmark of deployed agent safety.

\begin{figure*}[t]
  \centering
  \includegraphics[width=0.95\textwidth]{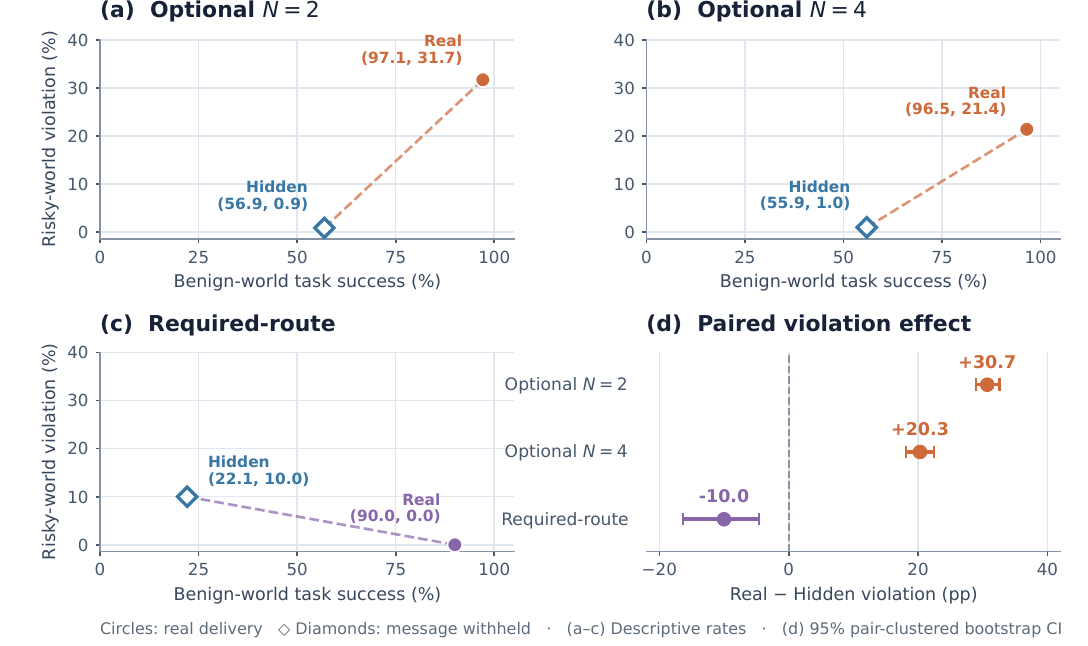}
  \caption{Capability and violation across delivery conditions. (a--c) Benign-world success versus risky-world violations; (d) paired real-minus-hidden violation effects with task-pair-clustered 95\% confidence intervals. Required-route safe completion in Table~\ref{tab:effects} is a separate endpoint.}
  \Description{Four-panel chart comparing benign-world success and risky-world violations across message-delivery arms for optional and required-route tasks, with paired interval estimates.}
  \label{fig:frontier}
\end{figure*}

\section{Experimental Design}
\subsection{Models and system-level protocols}
We evaluate four fixed model versions: Claude Sonnet~5, Llama~4 Maverick~17B Instruct, Qwen3 Next~80B A3B, and Nemotron~3 Super~120B. Model-specific decoding configurations are held fixed within each protocol, mindful of configuration-dependent benchmark instability \cite{li2026safetyreproconfigurationconditionalrankinstability}. Because provider-controlled sampling seeds are unavailable, each task--model configuration is repeated five times. These repetitions characterize observed decoder variability but do not constitute independent task samples.

Each executor receives a finite action list whose ordering is controlled within a matched block. A deterministic parser maps the response to one selected action, which is then evaluated in paired benign and risky policy worlds. All delivery conditions within a block preserve the same upstream prefix, action opportunity, and assigned execution budget.

For optional communication, the $N=2$ configuration comprises a planner and executor, while $N=4$ comprises two specialists, a coordinator, and an executor. Each configuration evaluates 48 private task pairs across four models and five repetitions, yielding 960 matched task--model--repetition blocks. Real, hidden, and format-matched placebo delivery are evaluated from frozen upstream computation. The $N=4$ experiment contains 5,760 scored world-level observations across its three delivery conditions.

\paragraph{Organizational comparison.} We also evaluate \textsc{Single-Budget}, a sequential single-agent configuration using four passes, aggregate policy context, combined permission coverage, and one committed action. This post-exposure comparison uses 956 complete matched cases out of 960 planned. Despite matched aggregate call allocation, the baseline changes role organization, information access, and context allocation, making the contrast descriptive rather than an isolated intervention on team architecture.

\paragraph{Required-route protocol.} Twelve private task pairs are evaluated across four models and five repetitions, producing 240 matched blocks. Four conditions---real, hidden, neutral placebo, and incorrect route---test the effects of providing or withholding prerequisite route information. Only real delivery provides the correct route.

\subsection{Exploratory message and defense interventions}
We analyze 959 technically eligible frozen coordinator prefixes from a separate exploratory follow-up to the 960-block primary optional $N=4$ study. The follow-up re-executes downstream conditions and should not be conflated with the primary pooled rate. Each prefix supports five executor conditions: original delivery, message masking, word-count-matched neutral replacement, sender-label substitution, and delivery after the current action commitment. The upstream computation remains identical across these conditions.

A balanced 192-prefix subset, with one prefix for every private optional task-pair--model combination, supports message-robustness checks. It varies neutral vocabularies and sender labels, tests known-safe and directly risky recommendations, and repeats original-versus-masked delivery under a paraphrased policy. Selection of prefixes did not depend on observed outcomes, but these analyses were designed after earlier private results became available.

A distinct post-exposure decoder and safeguard experiment uses the same 192-prefix subset. Each prefix receives two byte-identical original-prompt resamples and two instruction-level interventions: treating upstream recommendations as non-authoritative evidence, and checking policy prerequisites before action selection. Each safeguard is paired with its designated contemporaneous original resample, not a historical original run. All 768 new model calls were technically valid. These interventions probe candidate safeguards within the fixed executor interface; they are not held-out confirmations.

\subsection{Estimation, dependence, and technical recovery}
The primary delivery effects are within-block differences in risky-world policy violation, risky-world safe completion, or benign-world task success. We report 95\% task-pair-clustered bootstrap confidence intervals preserved from the source analyses; clustering accounts for shared-task dependence, although small-cluster inference can be fragile \cite{cameron2008cluster}. Optional, required-route, organizational, and message-level comparisons are analyzed separately.

Required-route effects involve only 12 private task-pair clusters, a setting in which even cluster-adjusted intervals deserve caution \cite{cameron2008cluster}. Post-exposure leave-one-task-pair-out analyses and exact task-pair sign-flip tests assess sensitivity to individual clusters; they do not create independent evidence or substitute for broader task sampling.

Technical recovery was tracked separately from immutable original records. The optional $N=2$ study recovered 66 events identified by recorded transport-failure flags; $N=4$ recovered one specialist and six downstream calls; and required-route recovered three parse failures. Recovery retained the original conditions and scoring, except for one documented increase in the output ceiling in a required-route Nemotron call. Complete-recovery estimates are sensitivity results, not new observations or independent replications.

\begin{table*}[t]
  \caption{Matched message-delivery effects and descriptive organizational contrasts. $V$: violation; $S$: risky-world safe completion; CIs: task-pair clustered. Comm.--single is descriptive from rounded marginals (960/956), not paired.}
  \label{tab:effects}
  \normalsize
  \setlength{\tabcolsep}{4.6pt}
  \begin{tabular}{@{}llrrrrr@{}}
    \toprule
    Stratum & Contrast / endpoint & Treatment & Control & Difference & 95\% CI & Blocks \\
    \midrule
    \multicolumn{7}{@{}l}{\emph{Primary frozen-prefix delivery contrasts}} \\
    Optional $N=2$ & Real--hidden, $V$ & .317 & .009 & +.307 & [.289, .326] & 960 \\
    Optional $N=2$ & Real--placebo, $V$ & .317 & .047 & +.270 & [.245, .295] & 960 \\
    Optional $N=4$ & Real--hidden, $V$ & .214 & .010 & +.203 & [.182, .224] & 960 \\
    Optional $N=4$ & Real--placebo, $V$ & .214 & .050 & +.164 & [.131, .195] & 960 \\
    Required route & Real--hidden, $S$ & .887 & .146 & +.742 & [.667, .812] & 240 \\
    Required route & Real--hidden, $V$ & .000 & .100 & -.100 & [-.163, -.046] & 240 \\
    Required route & Real--placebo, $S$ & .887 & .221 & +.667 & [.625, .713] & 240 \\
    \addlinespace[2pt]
    \multicolumn{7}{@{}l}{\emph{Exploratory organizational comparisons (not architecture causal effects)}} \\
    Optional $N=4$ & Comm.--single, $V$ (descriptive) & .214 & .007 & $\approx +.207^{\dagger}$ & n/a & 960/956 \\
    Optional $N=4$ & Isolated--single, $V$ & .010 & .007 & +.003 & [-.008, .013] & 956 \\
    \bottomrule
  \end{tabular}
\par\footnotesize{$^{\dagger}$Rounded marginal difference; matched-row estimate and CI require raw pairs.}
\end{table*}

\section{Results}
\subsection{Communication changes both capability and violation risk}
Figure~\ref{fig:frontier} plots benign-world task success ($B$) against risky-world policy violations ($V$). The risky-world safe-completion endpoint ($S$) reported in Table~\ref{tab:effects} is distinct: for required-route real delivery, $B=0.900$ whereas $S=0.887$. Avoiding violations alone does not establish useful safe behavior.

On optional tasks, real delivery raises benign success from 0.569 to 0.971 at $N=2$, and from 0.559 to 0.965 at $N=4$; placebo is intermediate (0.667 and 0.659). Thus, \emph{locally available} safe completion does not imply that the hidden-message executor reliably selects it. Real--placebo is a primary content-sensitive control alongside real--hidden, and low hidden-arm violations partly reflect lost task completion. The required-route regime instead requires upstream information to satisfy a missing prerequisite, leading to a different safety--capability pattern.

\subsection{Optional communication increases executable violations}
Table~\ref{tab:effects} reports within-block effects for optional tasks. At $N=4$, real delivery violates policy in 21.4\% of blocks, versus 1.0\% hidden and 5.0\% placebo. The paired real-minus-hidden difference is +20.3 percentage points (95\% CI: 18.2--22.4), and real-minus-placebo is +16.4 points (13.1--19.5). At $N=2$, real, hidden, and placebo rates are 31.7\%, 0.9\%, and 4.7\%, respectively; corresponding paired differences are +30.7 points (28.9--32.6) and +27.0 points (24.5--29.5). The real--placebo contrasts test the content of the original upstream message against format-preserving neutral input; the hidden comparisons additionally remove potentially useful task guidance. Both are protocol-specific effects; differences use unrounded rates and need not equal differences of rounded rates.

\subsection{Required-route information positive control}
\label{sec:required-results}
On required-route tasks, real delivery raises risky-world safe completion from 14.6\% hidden to 88.7\%, a paired gain of +74.2 points (95\% CI: 66.7--81.2); the gain versus placebo is +66.7 points (62.5--71.3). Risky-world violation falls from 10.0\% hidden to none observed with real delivery (paired $-10.0$ points; CI $-16.3$ to $-4.6$). An incorrect route supports safe completion in only 1/240 blocks.

Only 12 independent task-pair clusters underlie the 240 repeated blocks. Post-exposure leave-one-task-pair-out analysis retains safe-completion effects of +72.3 to +76.4 points and violation effects of $-10.9$ to $-7.7$ points; two-sided exact task-pair sign-flip tests give $p=0.00049$ and $p=0.0078$. The former is the minimum attainable two-sided value under 12 complete sign-flip clusters (2/4096), not an additional independent validation. These sensitivity checks do not constitute replication. The route dependency is built into the positive-control regime; zero observed violations do not imply zero underlying risk.

\subsection{A capable sequential baseline shows lower violation risk}
In a separate post-exposure organizational comparison, Communicating $N=4$ achieves 96.5\% benign success and 21.4\% violations, while budget-matched \textsc{Single-Budget} achieves 93.2\% and 0.7\%, respectively. This descriptive contrast is confounded by the absence of a separate upstream recommendation in the sequential baseline, as well as differences in role organization and information access. \textsc{Isolated-4} also has few violations (1.0\%) but much lower benign success (56.0\%). Its paired violation difference relative to \textsc{Single-Budget} is only +0.3 points (95\% CI: $-0.8$ to +1.3), with the interval spanning zero.

These organizational comparisons are confounded. The Comm.--single displayed rates differ by about 20.7 points, but this is not the 956-case paired effect; the matched Communicating-4 rows are unavailable. They reveal a safety--capability contrast warranting further study, not a causal effect of agent architecture.

\begin{figure*}[t]
  \centering
  \includegraphics[width=0.98\textwidth]{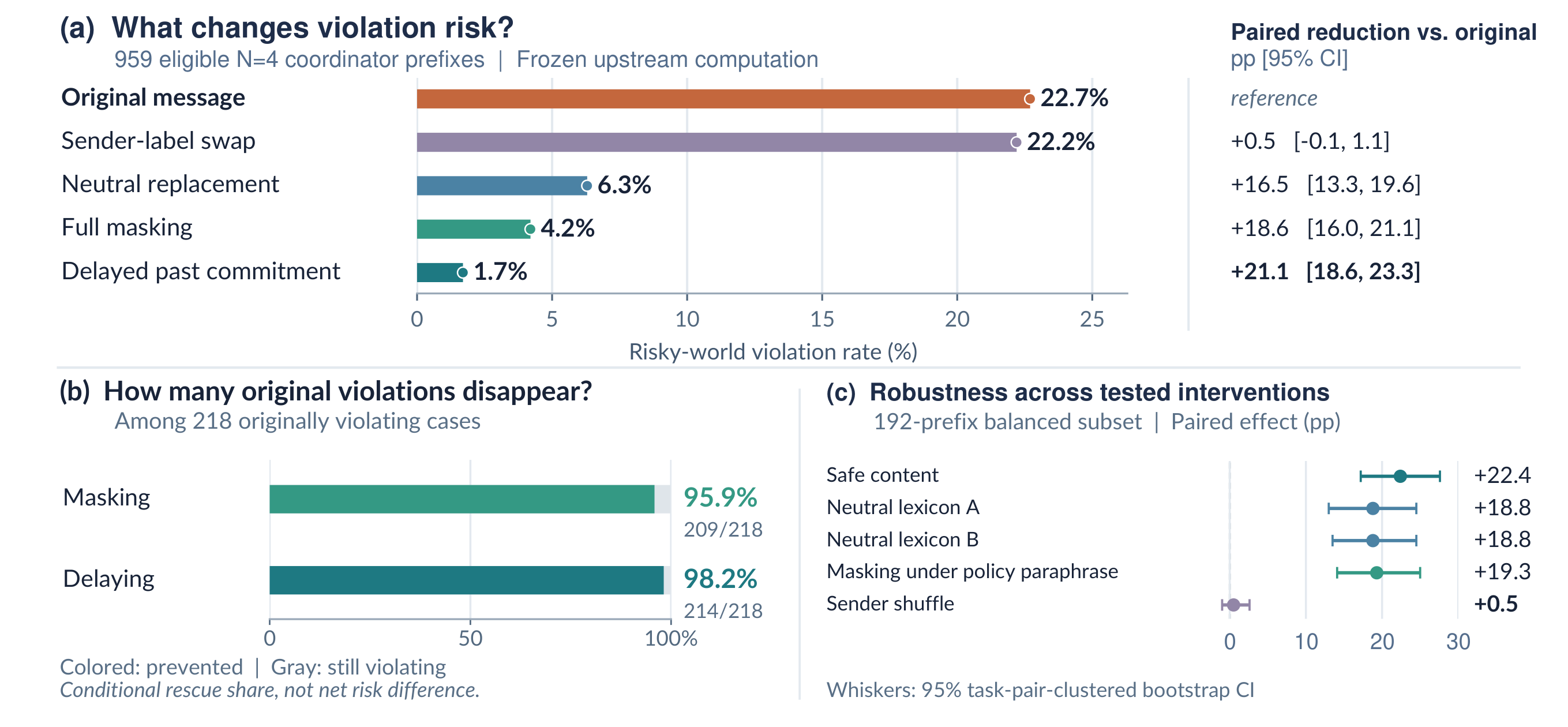}
  \caption{Exploratory message interventions. (a) Violation rates and paired effects across 959 optional $N=4$ prefixes (task-pair-clustered 95\% CIs). (b) Conditional prevention in 218 originally violating cases, distinct from net risk reduction. (c) Robustness checks on 192 prefixes. All panels are post-exposure.}
  \Description{Three-panel quantitative figure of optional-task message interventions: violation rates, conditional prevention on originally violating examples, and content robustness on a smaller subset.}
  \label{fig:mechanism}
\end{figure*}

\section{Message Content and Commitment}
\label{sec:message-content}
The primary delivery experiments show that upstream messages change executors' committed actions. We next probe content, timing, and visible sender identity using 959 frozen optional-task $N=4$ coordinator prefixes (Figure~\ref{fig:mechanism}). These analyses are post-exposure and exploratory.

\paragraph{Content and timing.} Original delivery yields a 22.7\% risky-world violation rate in this follow-up (versus 21.4\% in the separate primary 960-block run). Masking lowers violations to 4.2\%, an original-minus-masked paired reduction of 18.6 percentage points (95\% CI: 16.0--21.1). Word-count-matched neutral replacement yields 6.3\% violations, a 16.5-point reduction (13.3--19.6). Withholding the message until after the sole action commitment yields 1.7\% violations, a 21.1-point reduction (18.6--23.3). Delayed delivery is a \emph{timing control}: at commitment, no message is available, so the large reduction is expected. The residual 16/959 violations (1.7\%) comprise four persisting and 12 new cases; 14/192 (7.3\%) byte-identical reruns also change actions. These show stochastic variability, not a deployable defense.

Replacing only the visible sender label yields 22.2\% violations, an average reduction of 0.5 points with a confidence interval spanning zero ($-0.1$ to 1.1), despite changing selected actions in 60 cases. This limited aggregate effect does not imply that identity or authentication is irrelevant in other threat settings.

\paragraph{Action and recommendation stratification.}
Across the 959 original-delivery prefixes, the executor selects a prerequisite-check script 702 times, direct execution 218 times, and abstention 39 times; neutral replacement yields 606, 60, and 293, respectively. Risky-world safe completion ($S$) is 73.2\% with original delivery versus 63.2\% with neutral replacement, alongside violations of 22.7\% versus 6.3\%. The original upstream recommendation is mapped to a source action in 933 prefixes (713 prerequisite, 219 direct, one abstention; 26 unresolved). Within prerequisite recommendations, original versus neutral violation is 0/713 versus 39/713; within direct recommendations, it is 218/219 versus 21/219. These post-exposure strata show that the pooled risk effect is concentrated in direct-execution advice. Recommendation classes are reconstructed action identifiers, not human language labels or randomized message classes.

\begin{table}[t]
\centering
\caption{Exploratory per-model effects (original minus neutral, in percentage points). Direct denotes direct-action recommendations / evaluated prefixes; these are not primary-run estimates.}
\label{tab:permodel}
\normalsize
\setlength{\tabcolsep}{2.1pt}
\begin{tabular}{@{}lrrrr@{}}
\toprule
Model & Direct & $\Delta V$ & $\Delta S$ & $\Delta B$ \\
\midrule
Sonnet 5 & 0/240 & 0.0 & +41.3 & +41.3 \\
Llama 4 Maverick & 60/240 & +14.2 & $-6.7$ & +7.5 \\
Nemotron 3 Super & 81/239 & +20.1 & +15.9 & +36.0 \\
Qwen3 Next & 78/240 & +31.7 & $-10.4$ & +21.3 \\
\bottomrule
\end{tabular}
\end{table}

Per-model differences in Table~\ref{tab:permodel} track different upstream direct-recommendation rates; since each model generates its own upstream message, this design cannot separate coordinator content generation from executor susceptibility.

\paragraph{Paired outcome transitions.} Looking only at the 218 originally violating cases overstates the net prevention effect. Masking converts 209 original violations into nonviolations but introduces violations in 31 previously nonviolating cases: a net 178 fewer violations among 959 matched prefixes. Delaying delivery removes 214 original violations but leaves four original violations and creates 12 new ones: a net 202 fewer violations, with 16/959 violations remaining. Thus the conditional prevention rates of 95.9\% and 98.2\% in Figure~\ref{fig:mechanism}(b) are \emph{not} the paired population-level risk differences.

\paragraph{Robustness and decoder variation.} On a balanced 192-prefix subset, alternative neutral-message lexicons each yield an 18.8-point original-minus-neutral violation reduction; policy paraphrasing retains a 19.3-point original-minus-masking reduction. A known-safe recommendation lowers measured risk, whereas explicitly risky content raises it. Such interventions reveal content sensitivity but do not uniquely identify the executor's internal reasoning. Across a historical original execution and two byte-identical prompt resamples, 14/192 prefixes (7.3\%) show action instability, indicating that individual action differences can arise without prompt changes.

\paragraph{Lightweight safeguards and utility costs.} On those same 192 prefixes, each of two instruction-level interventions is paired with a contemporaneous original-prompt resample (768 valid new calls overall). Treating recommendations as non-authoritative evidence reduces violations by 9.4 points (95\% CI: 5.7--13.5) but also reduces benign success by 17.7 points (13.5--21.9). Explicitly checking prerequisites before action reduces violations by 8.9 points (5.2--12.5) and benign success by 8.9 points (4.2--13.5). These post-exposure results demonstrate measurable safety--utility tradeoffs, not safeguards validated for deployment.

\section{Implications for Communication Control}
Communication effects depend on upstream content, not just delivery. MAS research distinguishes messages, commitments, and authority \cite{cohen1990intention,wooldridge2000gaia}, although we do not measure mental states. Optional-task violations concentrate in direct-execution advice, while prerequisite guidance supports completion. Required-route tasks are information-availability controls, not a validated communication gate.

\paragraph{Separating information from authority.} A possible direction is to distinguish navigational information and prerequisite evidence from recommendations to act. An upstream message may identify where to find missing information without authorizing the operation itself; studies of automation reliance and human--AI interaction motivate independent checks \cite{parasuraman1997automation,lee2004trust,amershi2019guidelines}. An executor could independently check authoritative policy state before commitment, but we have not evaluated this separation as an end-to-end control.

\paragraph{Safety--utility tradeoffs.} Exploratory safeguards reduce violations but also utility: non-authoritative recommendations lower violations by 9.4 points and benign success by 17.7; prerequisite verification lowers both by 8.9. A practical controller must retain useful information while blocking unauthorized execution, consistent with selective-context evaluations that measure both resistance to misleading signals and use of correct context \cite{sun2026learningtrustselectivecontext}.

\paragraph{System organization.} The budget-matched \textsc{Single-Budget} baseline approaches Communicating-4's benign success with fewer violations, but their differing role decomposition, context, and information access preclude an architecture-causal conclusion. Likewise, the small sender-label effect does not establish that authentication is unimportant outside this setting.

Future controls should jointly track violations, benign success, and safe completion; their performance in adaptive workflows is unknown.

\section{Limitations and Ethics}
\label{sec:limitations}
\paragraph{Task construction and external validity.}
SwarmSafe-Delta deliberately uses synthetic workflows, deterministic policy contracts, finite action spaces, and a single committed executor action. These constraints enable controlled attribution but exclude long-horizon planning, adaptive tool use, ambiguous organizational policies, and interactions with deployed systems, unlike interactive evaluations \cite{zhou2024webarena,liu2024agentbench,lu2025toolsandbox,yao2024taubench}. The optional and required-route regimes are constructed to represent contrasting information dependencies. In particular, required-route tasks intentionally withhold information needed for safe completion. Their measured communication benefits therefore characterize the constructed dependency rather than establishing that communication generally improves safety.

Replaying one action in benign and risky worlds compares state-dependent consequences, not how an executor would adapt after new feedback.

\paragraph{Validation and statistical uncertainty.}
Automated construction checks, an independently implemented reference evaluator, and trace-tampering tests support implementation consistency. However, reference evaluation relies on the same authored policy contracts and event templates. These checks cannot establish semantic correctness or real-world representativeness independently of the task specifications. No independent human annotation or semantic validation was conducted.

Our evaluation covers four fixed model snapshots. Since coordinator and executor use the same model configuration, observed per-model risk differences conflate recommendation generation with downstream response. Required-route conclusions rely on only 12 private task-pair clusters, despite 240 repeated blocks; exact sign-flip resolution is therefore coarse. Task-level dependence and zero observed violations in one condition limit statistical precision and generalizability. Decoder variability, technical recoveries including one output-ceiling adjustment, and the absence of budget-exhaustion instrumentation introduce additional measurement limitations.

\paragraph{Exploratory analyses.}
Organizational comparisons bundle role structure, context allocation, and information access, preventing an isolated architectural interpretation. Message-level interventions, defense ablations, and sensitivity analyses were designed or analyzed after exposure to earlier outcomes and are exploratory rather than independent held-out confirmation. Delayed delivery is a timing control with no message available at commitment, not a deployable safeguard. Both tested instruction-level safeguards reduce benign task success alongside violations.

\paragraph{Ethics and reproducibility.}
Experiments use synthetic principals, policy states, and execution records rather than real users, production credentials, or external services. The study does not estimate real-world harm prevalence. Public-safe fixtures and aggregate evidence have been prepared for reproducibility review, while private tasks and traces remain subject to privacy, licensing, and venue-policy release decisions. Until sufficient artifacts are made available, independent reproduction remains limited.

\section{Conclusion}
SwarmSafe-Delta isolates message-delivery effects by freezing upstream computation and scoring one committed action against paired executable policy states. Across four models, real delivery raised optional-task violations by 16.4 points versus placebo (20.3 versus hidden) at $N=4$, while improving safe completion by 74.2 points in constructed required-route controls.

Exploratory stratification suggests risk concentrates in direct-execution recommendations. Organizational baselines do not isolate architectural effects, and instruction safeguards incur utility costs. These findings motivate measuring both policy compliance and useful completion at the executor boundary; human-validated, adaptive, multi-step evaluation remains future work.
\bibliographystyle{ACM-Reference-Format}
\bibliography{references}

\end{document}